\documentclass[aps,prl,twocolumn,showpacs]{revtex4}

\usepackage{graphicx}
\DeclareGraphicsExtensions{.png,.jpg,.eps}

\usepackage{xcolor}
\usepackage{amsmath}

\begin{document}

\title{Van der Waals spin valves }

\author{ C. Cardoso, D. Soriano, N. A. Garc\'ia-Mart\'inez,  J. Fern\'andez-Rossier \footnote{On leave from Departamento de F\'isica Aplicada, Universidad de Alicante,  Spain} }
\affiliation{QuantaLab, International Iberian Nanotechnology Laboratory (INL),
Av. Mestre Jos\'e Veiga, 4715-330 Braga, Portugal
}

\date{\today}

\begin{abstract}
%
We propose  spin valves  where  a 2D  non-magnetic conductor  is intercalated between two ferromagnetic insulating layers.  In this setup, the relative orientation of the magnetizations of the insulating layers can have  
  a strong impact on the in-plane conductivity of the 2D conductor.
We first show this for a
 graphene bilayer, described with a tight-binding model,  placed between two ferromagnetic  insulators. In 
 the anti-parallel configuration, a band  gap opens at the Dirac point, whereas in the 
  parallel configuration, the  graphene bilayer remains conducting.
  We then compute the electronic structure of graphene bilayer  placed between two monolayers of the ferromagnetic insulator  CrI$_3$, using density functional theory. Consistent with the model, we  find that a gap opens at the Dirac point only in the antiparallel configuration.

\end{abstract}

\maketitle


The controlled fabrication of layered structures combining ferromagnetic conductors and non magnetic materials, thin enough as to preserve spin polarization, made possible the discovery of fundamental spin dependent transport phenomena, such as Giant Magnetoresistance ~\cite{baibich1988,binasch1989} and  tunnel Magnetoresistance \cite{julliere1975,moodera1995}. These developments led to the concept of spin valve, a
structure whose conductivity    is modulated by the relative orientation of  two ferromagnetic electrodes  \cite{dieny1991} and, altogether,  set the foundations of spintronics.

The study of the so called Van der Waals heterostructures~\cite{Geim2013,novoselov2016},  metamaterials obtained by vertical stacking of 2D crystals, is a very fertile area of research.  Using relatively simple fabrication methods,  they allow the study of structures with tailored  electronic properties that combine a variety of 2D materials, including insulators  (h-BN),  semiconductors (MoS$_2$), conductors (graphene) and superconductors  (NbSe$_2$).   
The recent discovery of 2D crystals with magnetic order \cite{Wang2016,Lee2016,Gong2017,Huang2017} adds
both  ferromagnetic and antiferromagnetic insulators to this list.   For instance,   Van der Waals devices  incorporating atomically thin layers  of  the  ferromagnetic insulator CrI$_3$,  have being reported\cite{klein2018,song2018,wang2018,Jiang2018,Huang2018,Jiang2018b}. These findings pave the way to  Van der Waals spintronics with new types of spin dependent transport phenomena.
Here
we propose a new class of spin valve that  takes advantage of the   spin proximity effect,  {\em i.e.},  the spin polarization of the surface electrons of a  non-magnetic material adjacent to a ferromagnet. 
 The  proposed system,  depicted in Figure 1, consists of a 
2D conducting  crystal sandwiched between two insulating ferromagnetic layers.
If the magnetizations of the two proximity layers are anti-parallel (AF),  the  spin proximity effects have opposite sign at both sides of the 2D crystal.   In contrast, for the parallel state (FM),  the top and bottom  proximity effects add up.  As we show below, this difference has a strong influence in the in-plane conductance of the non-magnetic conductor,  and in some instances leads to a   {\it conductor to insulator transition} in the 2D crystal. 
This strong influence of spin proximity effect in a 2D crystals contrast with the case of 3D materials, for proximity effects are   constrained to their surface. 

Our proposal is different from lateral graphene  spin valves~\cite{tombros2007,han2014}, where large areas of the graphene are not in contact with ferromagnetic electrodes, and  is also  different from spin-filter tunnel junctions, where the magnetic insulators act as barrier materials for vertical transport \cite{Miao2009,klein2018,song2018,wang2018}.   The proposed  spin valve resembles the early current in plane structures where giant magnetoresistance was discovered~\cite{baibich1988,binasch1989}, with the obvious difference that the magnetic layers are insulating in the Van der Waals spin valves.

\begin{figure}[h!]
 \centering
    \includegraphics[width=0.4\textwidth]{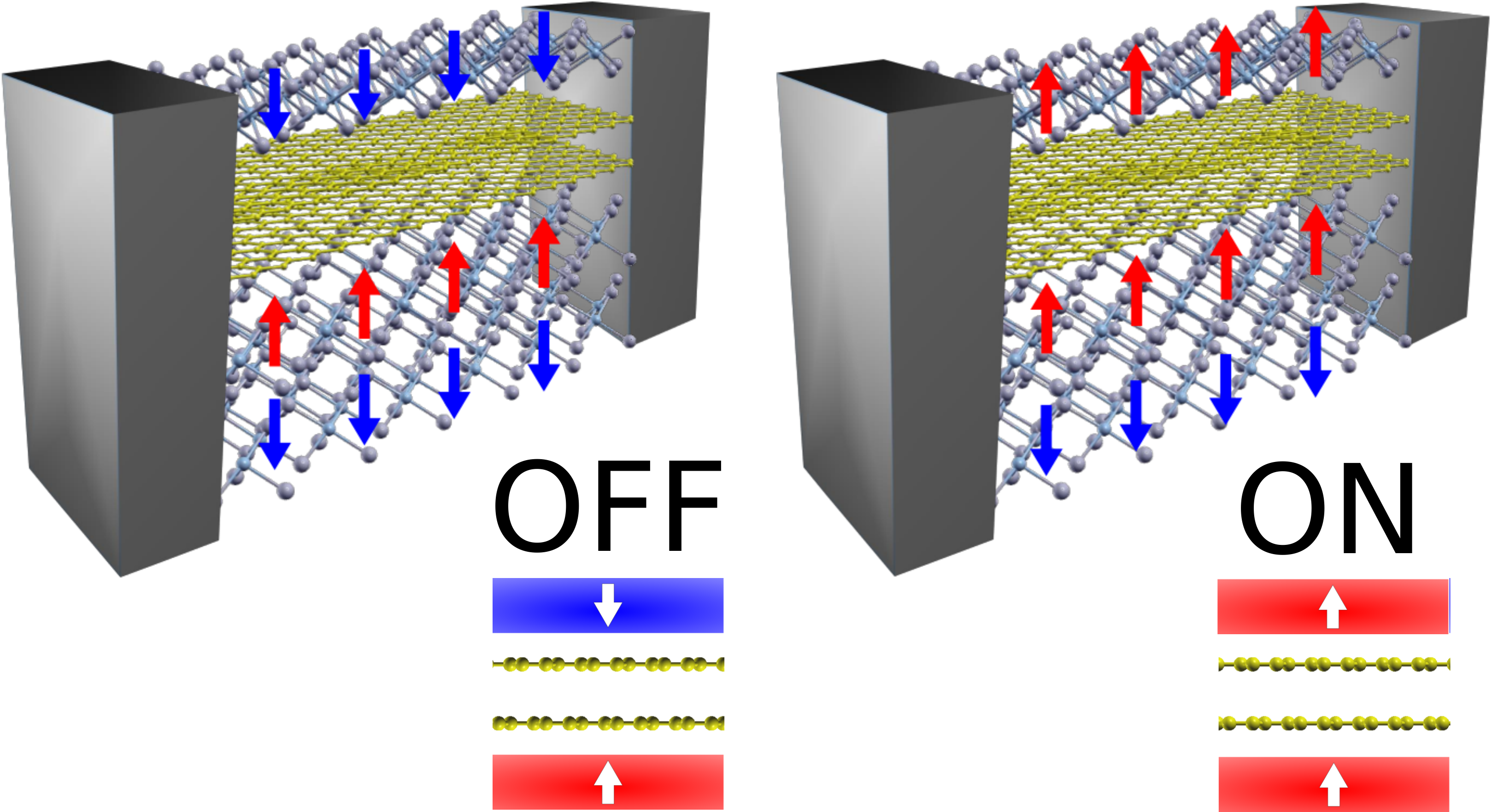}
\caption{  Van der Waals spin valve, where a conducting 2D crystal  is sandwhiched between 2 insulating ferromagnets. Lateral 
contacts can drive in-plane current.   
The magnetization of the bottom ferromagnetic layer  is pinned, whereas the top layer can switch, resulting in two configurations, 
(a) antiparallel (AF)
and (b)  parallel (FM),  with very different in-plane conductance. 
  }
\label{fig1}
\end{figure}

 We  first illustrate the concept of Van der Waals (VdW) spin valves  considering the case when the central conductor is a graphene bilayer. 
We use  the standard  tight-binding  model for the  graphene bilayer\cite{guinea2006}.   
Spin proximity effect is considered\cite{phong2017} by including a spin  dependent potential $\Delta$ whose sign can be different in the top and bottom graphene layers.  Previous density functional theory (DFT) calculations for monolayer graphene deposited on different ferromagnetic insulators, such as EuO\cite{yang2013,hallal2017},  EuS \cite{hallal2017}, YIG \cite{hallal2017}, justify this model.  We assume that the magnetization of both top and  bottom layers  lie on the same axis. We also assume that  the bottom magnetization does not change, resulting in fixed  spin-dependent  potential $\sigma \Delta$, where $\sigma$ is the spin projection  along the magnetization axis.    The top layer spin dependent potential 
is given by $\eta \sigma \Delta$, where $\eta=\pm 1$ describes the magnetization  orientation of the top magnetic layer, relative to the bottom layer.  Thus, $\eta=+1$ describes the parallel (FM) orientation and $\eta=-1$ the antiparallel (AF) case.

\begin{figure}[h!]
 \centering
 \includegraphics[width=0.45\textwidth]{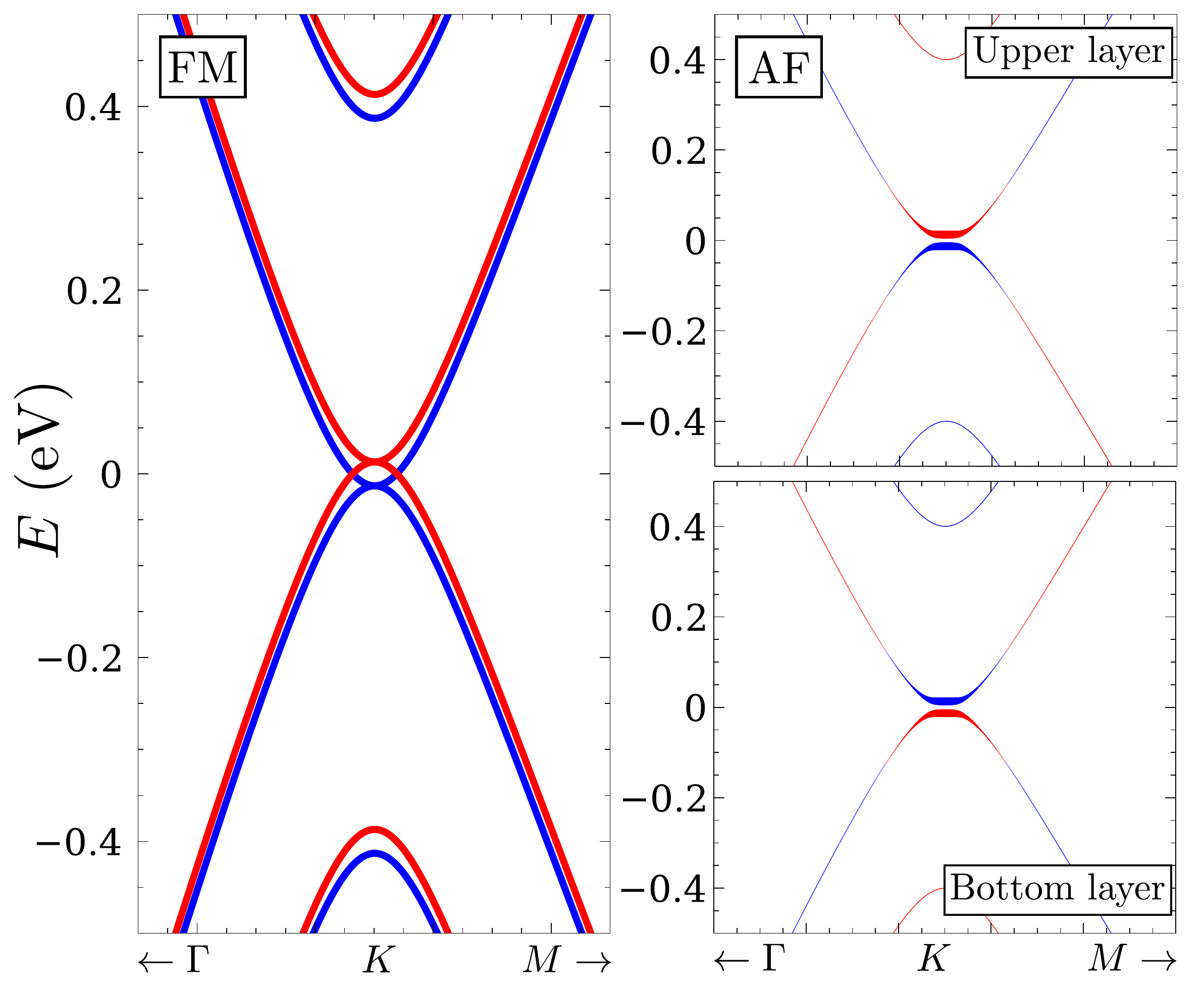}
\caption{Bands structure computed with model Hamiltonian for the FM (left) and  AF (right) configurations, 
Red and blue stand for the different spin states. 
 For the AF state, the bands have been projected in the upper and lower layer of the graphene bilayer, with the size of the dots proportional to the layer polarization, revealing that the top of the valence band and bottom of conduction band are spin polarized in each of the layers.
}
\label{fig2}
\end{figure}

Represented in the basis $A_1,B_1, A_2, B_2$, where $A$ and $B$ correspond to the two triangular sublattices, and the subindices 1 and 2  label the top and bottom layers, respectively, the Bloch Hamiltonian for spin $\sigma$ states reads:
\begin{eqnarray}
{\cal H}_{\sigma}(\vec{k})=\left(
\begin{array}{cccc}
\eta\sigma \frac{\Delta}{2} & f(\vec{k}) & 0& 0 \\
f^*(\vec{k}) & \eta\sigma \frac{\Delta}{2} & \gamma  & 0 \\
0 & \gamma  &  \sigma \frac{\Delta}{2} &  f(\vec{k})\\
 0 & 0 &  f(\vec{k})^* & \sigma \frac{\Delta}{2}
\end{array}
\right)
\label{hmodel}
\end{eqnarray}
where $f(\vec{k})=t\left(1+e^{i\vec{k}\cdot\vec{a}_1}+e^{i\vec{k}\cdot\vec{a}_2}\right)$ and $\gamma$ describe the intralayer and interlayer hopping matrix elements, respectively. The resulting spin resolved energy bands, in the neighbourhood of the Dirac point, are shown in fig.~\ref{fig2} for the two states of the spin valve, $\eta=\pm 1$.
For the FM alignment ($\eta=+1$), the  graphene bilayer presents spin-split bands, and remains in a  conducting state,{\it  i.e.}, with a finite density of states at the Fermi energy.  In contrast, for the  AF case ($\eta=-1$), a band-gap  opens up at the Dirac point.
Thus,  depending on the relative alignment of the top and bottom insulating ferromagnets, 
  the graphene bilayer  spin valve can be either a conductor,  for the FM alignment,  or a gapped system with depleted conductance, when  the Fermi energy is set at  the Dirac point.  Within this model, both the band-gap in the AF alignment  and the spin splitting in the FM  alignment are given by $\Delta$. 

We now address the origin of the gap opening in  graphene bilayer AF alignment of the VdW spin valve.
For each spin channel, 
the  Hamiltonian  (\ref{hmodel}) in the AF alignment  is identical to the model of graphene bilayer with 
an off-plane electric field, that is known to open up a gap in the band structure~\cite{ohta2006,mccann2006}. 
Interestingly, in the spin-valve, the sign of the effective electric field is opposite for opposite spins, $E_{\rm eff}\propto \sigma \Delta$. 
The spin projection of the AF bands over top and bottom layers, shown in Fig. \ref{fig2}, clearly shows the presence of a spin dipole\cite{JFR2008}: for a given spin,  there is a charge imbalance driven by the exchange with the magnetic layers, that is compensated exactly by the opposite spin.

In a graphene bilayer, 
the gap opened by an electric field 
is known to have a valley dependent Chern number ${\cal C}= K {\rm sgn}(E)$, where $K=\pm 1$ labels the valleys~\cite{martin2008,pablo2009}. In the case of the spin valve in the AF state, this leads  to  Chern numbers that are both spin and valley dependent:
\begin{equation}
{\cal C}= K \sigma=\pm 1
\label{chern}
\end{equation}

\begin{figure}[hbt!]
 \centering
  \includegraphics[width=0.4\textwidth,angle=0]{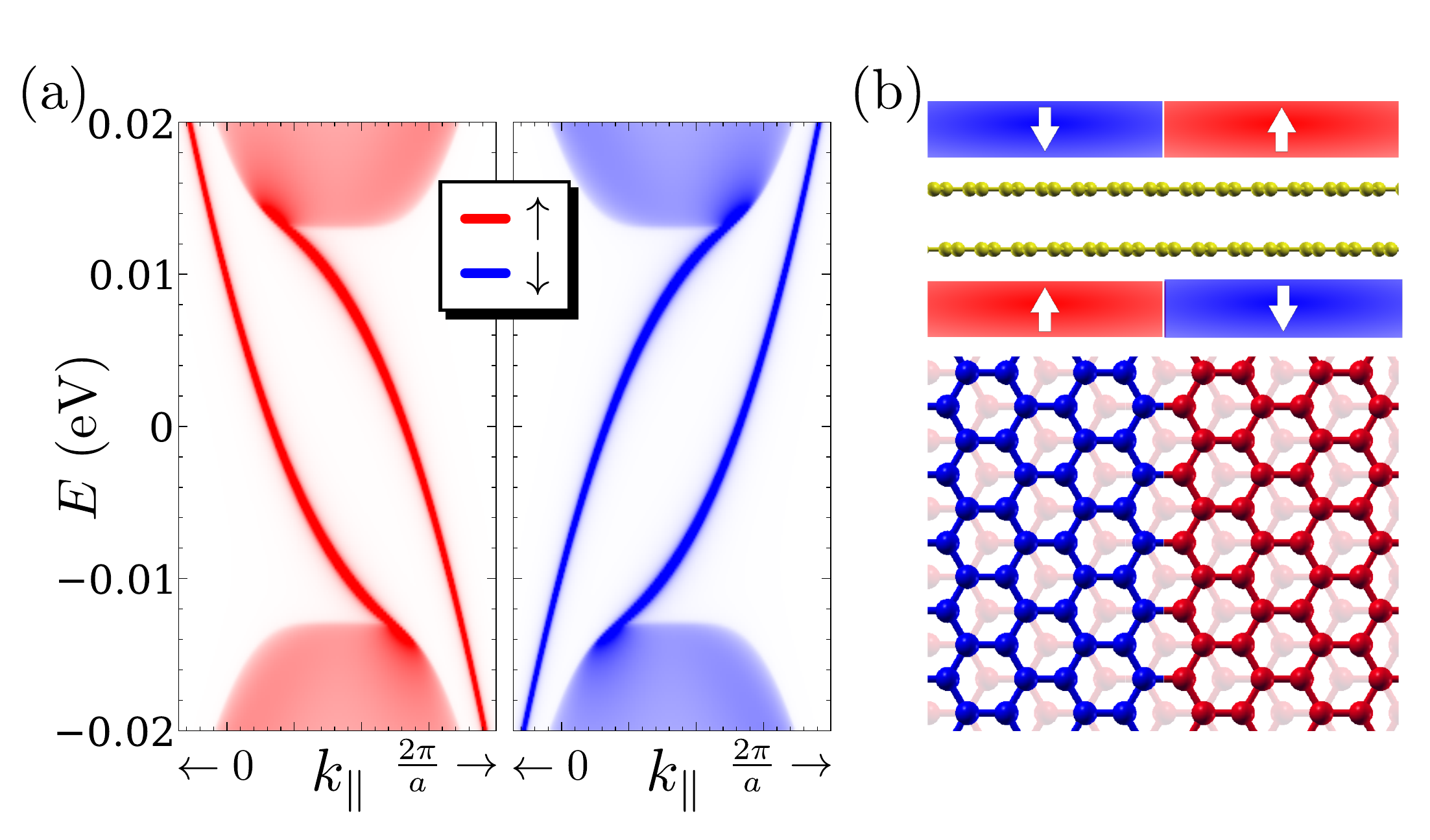}
\caption{ Chiral in-gap states, for a given valley, computed for the domain wall between two  insulating 
antiferromagnetic domains with opposite spin orientation.  The velocity of the bands changes sign in the opposite valley.  }
\label{fig3}
\end{figure}

Eq.~\eqref{chern} permits to anticipate~\cite{martin2008,pablo2009}  the emergence of spin-locked chiral one dimensional in gap states in domain walls separating two  antiferromagnetic 
domains with opposite magnetizations  (see Fig \ref{fig3}). 
In order to verify this, we compute the momentum resolved density of states
of a domain wall along the zigzag direction 
as ilustrated in fig.~\ref{fig3}. The domain wall is assumed to be abrupt, preserving spin collinearity. The calculation is done for a system with translational invariance along the wall direction, and embedded between two semi-infinite gapped graphene bilayer planes, using a Green function technique~\cite{Lado2015}.
Both domains are insulating, but at each valley the Chern number is opposite for a given spin direction. Thus,  a domain wall along the zigzag direction, that preserves the valley, features two chiral 1D in-gap states per valley and per spin (see fig.~\ref{fig3}). It is interesting to note that, for a given valley, the states are spin chiral and therefore back-scattering requires either spin-mixing, or inter-valley scattering. A non collinear domain wall might result in spin mixing.

We now consider a possible physical realization of the Van der Waals spin valve, based on a graphene bilayer, that is feasible within the experimental state of the art. For that matter, we choose  CrI$_3$  monolayers as the insulating ferromagnet. It was recently shown that CrI$_3$ monolayers  remain ferromagnetic up to 45K ~\cite{Huang2017}. 
In addition  CrI$_3$ preserves its magnetic properties even when  deposited on graphite~\cite{Huang2017} or encapsulated between graphite electrodes~\cite{klein2018,song2018,wang2018}.   Moreover, 
 the spin proximity effect between CrI$_3$ and two dimensional WSe$_2$ has been demonstrated experimentally \cite{zhong2017}.  
 
The DFT calculations were performed using Quantum ESPRESSO~\cite{Giannozzi_JPhysConMat_2009} 
with PBE exchange-correlation potential~\cite{Perdew_JCP_1996} and PAW pseudopotentials~\cite{PhysRevB.50.17953,kucukbenli2014projector} and including van der Waals interactions within the semiempirical method of Grimme (DFT-D2)~\cite{Grimme_JCompChem_2006}.
Spin orbit interactions, known to be important to determine the magnetic anisotropy  of CrI$_3$~\cite{Lado2017}, are not included in the calculation,  as they are not expected to induce qualitative change in the spin proximity effect discussed here.

\begin{figure}[hbt!]
 \centering
 \includegraphics[width=0.48\textwidth]{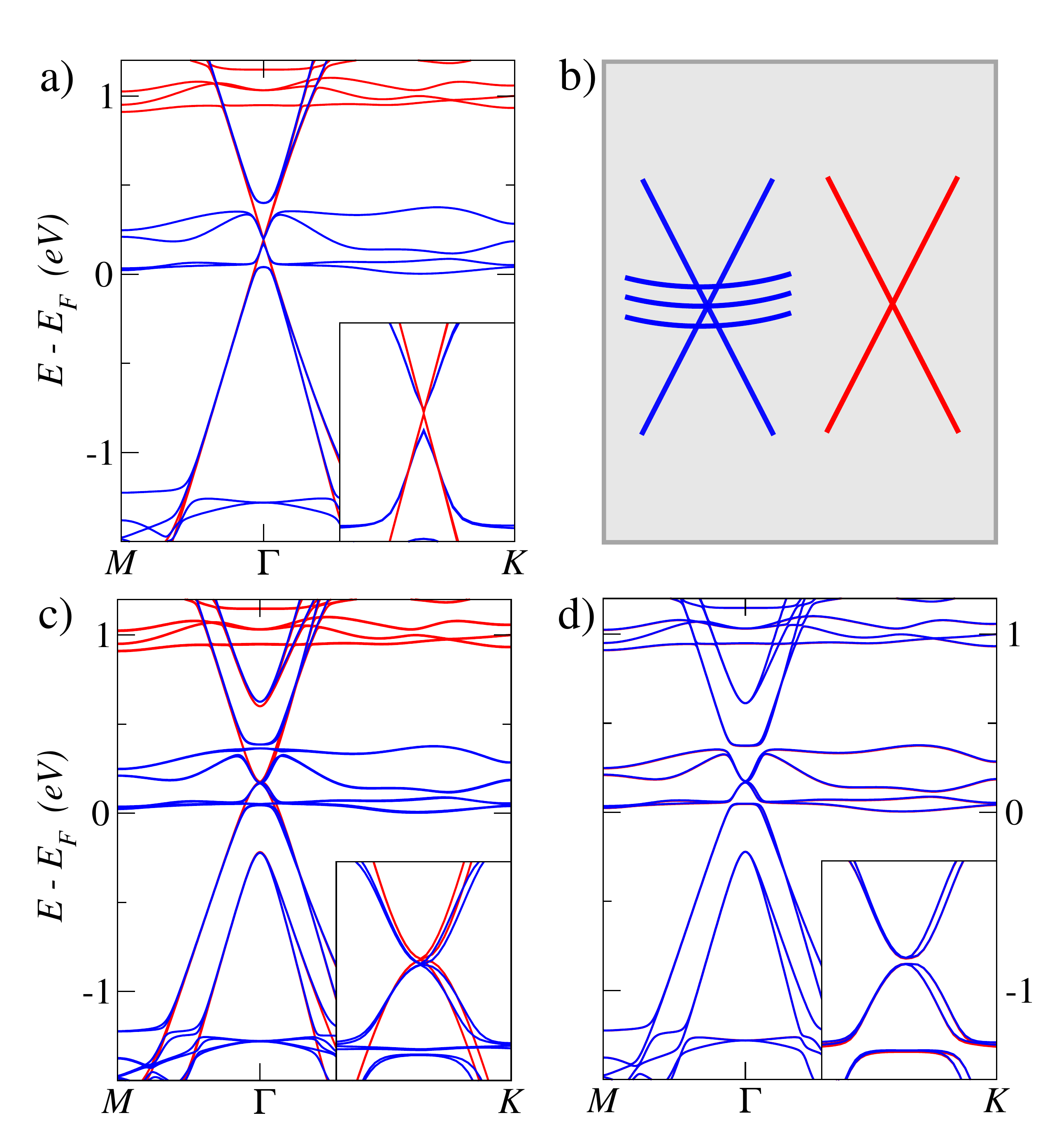}
\caption{(a) Band structure,   for the graphene monolayer on top of CrI$_3$ monolayer.  
(b) Scheme of energy bands for  graphene monolayer on CrI$_3$.   (c) and (d)   Band structure for grahene bilayer placed between ferromagnetic monolayers of CrI$_3$ with FM (c) and AF (d) relative alignemnent.  The insets show a zoom around the Dirac point, showing a spin splitting in the FM alignment and a gap for the AF case,  both taking the same value  of 7  meV. }
\label{fig4}
\end{figure}

We now discuss our results  both  for monolayer CrI3 /monolayer graphene  as well as graphene bilayer in between two monolayers of CrI$_3$.    The former permits to rationalize the results of the bilayer.
For the bilayer, we use
 a unit 
cell with  4 layers, CrI$_3$/graphene bilayer/CrI$_3$, and two different geometries: 1) a free standing multilayer and 2) a  superlattice with periodic boundary conditions along the off-plane direction. 
In the case of the superlattice with periodic boundary conditions, we considered the lattice parameter of graphene for the plane and varied the $c$ lattice parameter in order to optimize the interlayer distance by minimizing the total energy. 
For the free standing multilayer, we used a lattice parameter along $z$ that ensures the absence of interaction between the replicas of the system.  
In all  cases the unit cell contains a 3$\times$3 supercell with 18 carbon atoms per graphene layer, 2 chromium and 6 iodine atoms per CrI$_3$ layer. Thus,  for the sandwiched graphene bilayer, the unit cell has a total of 52 atoms. 
The calculations show that the spin  of Cr atoms is $S=3/2$, that are hosted by the $t_{2g}$ bands.

We first discuss the results for the monolayer graphene on top of CrI$_3$, shown in (\ref{fig3})(a). Our results are in line with previous DFT calculations for this system~\cite{zhang2017}. 
With the exception of some anti-crossings, the energy bands are an overlay 
of those of  the decoupled monolayers,   
as expected in a Van der Waals structure.  Occupied bands, way below the Fermi energy $E_F$, are made of iodine $p$ states and the spin majority $t_{2g}$ states of Cr.  Empty bands, high above $E_F$, are made of spin minority  $t_{2g}$ states of Cr. 
In the 2 eV window around the Fermi energy, the  bands  are those of the graphene Dirac cones and the  spin majority $e_g$ states.  These four  bands (coming from 2 Cr atoms in the unit cell)  are narrow,    lie almost completely above the Fermi energy, and hybdridize with the  graphene Dirac cones in the majority spin channel close to the Dirac point.  In contrast,  the minority spin Dirac cone remains intact (see cartoon in (\ref{fig4})(b).  Therefore,  for one spin channel  the Dirac electrons barely notice the presence of the CrI$_3$, for the other spin channel, there is a strong hybdrization with a narrow band. 

The bands structure of the CrI$_3$/graphene bilayer/CrI$_3$ are shown in figure (\ref{fig3}), both for the FM (left) and AF (right) configurations.  Both of them show the graphene bilayer bands and the $e_g$ bands. 
For the FM case,  the shape of the graphene bilayer bands is preserved in one spin channel (the minority spin), but a strong hybridization opens up a gap in the majority spin channel, slightly above the Fermi energy, which lies below the Dirac point. For the AF configuration, both spin channels of the graphene bilayer become hybdridized with the narrow $e_g$ bands of CrI$_3$.  So,  from that point of view alone,  we expect that the in-plane conductance is much larger in the FM alignment than the  AF one. 

We now turn our attention to the  
states around the Dirac energy, where the conduction and valence parabolic bands of freestanding graphene bilayer meet. 
  For the AF configuration, a band-gap splits the electron and hole parabolic of  graphene bilayer (see inset of Fig. (\ref{fig4}d)), with a gap of   7 $meV$. For the FM configuration, there is a spin splitting of  the bands near the Dirac point,  whose magnitude   is, interestingly, the same,   than the AF gap.  So, in that regard,  the DFT results for the bilayer graphene placed between two CrI$_3$ layers are in line with the toy model. 
     However, there is electron  transfer from graphene to CrI$_3$, so that the Fermi energy does not lie at the Dirac point. It must be noticed that, even if there is charge transfer,  we expect a very small   in-plane conductance in the CrI$_3$, on account of the very small dispersion of the occupied states.

Thus, our  DFT calculations  strongly suggest that  a graphene {\em bilayer } encapsulated between two layers of CrI$_3$ will present a strong spin-valve effect, with the FM alignment having a larger in-plane conductance, on account of the fact that bands in one spin-channel are decoupled from the non-dispersive CrI$_3$ bands, in contrast with the AF, where both spin channels are hybdridized.  Interestingly,  the same mechanism should also apply for the graphene {\em monolayer}.   In addition,  for the graphene bilayer,  application of a gate voltage could set  the Fermi energy at the Dirac point, resulting in a  conductor to insulator transition  driven by the alignment of the magnetizations. 
 
 The control of the relative orientation of the magnetization of the layers could be done by application of a magnetic field, provided that two conditions are met. First, the interlayer coupling should be smaller than the Zeeman coupling.   Interlayer coupling of CrI$_3$ bilayers, without graphene in the middle, meets this demand.  The presence of the graphene bilayer should significantly reduce the interlayer coupling.  Accordingly, our
DFT calculations yield  $J< 10\mu$eV per unit cell.  Second, the switching field of top and bottom layers should be different.  A very natural way to achieve that is to   pin the magnetization of the bottom layer. This could be done, for instance, using    a  bilayer of CrI$_3$  for which antiferromagnetic interlayer coupling has been reported \cite{Huang2017,klein2018,Jiang2018,Huang2018,Jiang2018b} 

The concept of Van der Waals spin valve goes beyond the case of the graphene bilayer.
  For instance,
the tunable spin proximity effect can drive a metal insulator transition in a  graphene monolayer   in the Quantum Hall regime.  At half filling,   high quality graphene quantum Hall bars  are often  insulating. Application of a  strong in-plane magnetic field can induce a Quantum Hall ferromagnet that has a finite edge conductance~\cite{Young2014,Lado2015}.   In a spin valve, such transition could be promoted by
 spin proximity effect, rather than Zeeman interaction. 
 Another  possibility are Van der Waals spin valves with a superconducting middle layer,
such as NbSe$_2$.
In the FM state, spin proximity effect can kill superconductivity, that would only survive in the AF state.  
Such a transition has been observed in superconducting thin films sandwiched between bulk ferromagnetic insulators \cite{li2013}. The concept of Van der Waals spin valve can be extended to the control of optical properties. For instance, spin proximity might control  whether  dark or bright excitons of the middle layer are the ground state of the 2D semiconductor.

In conclusion, we have proposed a new type of spin valve where a non-magnetic 2D crystal is sandwiched between two ferromagnetic insulators and its in-plane conductance is controlled by spin proximity effect.  We hope that our work will motivate the experimental exploration of  Van der Waals current in plane spin valves,  including  the  of other  materials in the non-magnetic layers, such as  superconductors,  as well as other magnetic layers, such as  bulk ferrromagnets,  for which the spin valve effect proposed here should also work.

We acknowledge J. L. Lado for fruitful discussions and technical assistance in the calculations.
We  acknowledge financial support from FEDER project NORTE-01-0145-FEDER-000019, 
the Marie Curie Nano TRAIN for Growth Cofund program at INL and  FCT for the P2020-PTDC/FIS-NAN/3668/2014 project.
We acknowledge Efr\'en Navarro-Morata for fruitful discussions.


\begin{thebibliography}{40}
\expandafter\ifx\csname natexlab\endcsname\relax\def\natexlab#1{#1}\fi
\expandafter\ifx\csname bibnamefont\endcsname\relax
  \def\bibnamefont#1{#1}\fi
\expandafter\ifx\csname bibfnamefont\endcsname\relax
  \def\bibfnamefont#1{#1}\fi
\expandafter\ifx\csname citenamefont\endcsname\relax
  \def\citenamefont#1{#1}\fi
\expandafter\ifx\csname url\endcsname\relax
  \def\url#1{\texttt{#1}}\fi
\expandafter\ifx\csname urlprefix\endcsname\relax\def\urlprefix{URL }\fi
\providecommand{\bibinfo}[2]{#2}
\providecommand{\eprint}[2][]{\url{#2}}

\bibitem[{\citenamefont{Baibich et~al.}(1988)\citenamefont{Baibich, Broto,
  Fert, Van~Dau, Petroff, Etienne, Creuzet, Friederich, and
  Chazelas}}]{baibich1988}
\bibinfo{author}{\bibfnamefont{M.~N.} \bibnamefont{Baibich}},
  \bibinfo{author}{\bibfnamefont{J.~M.} \bibnamefont{Broto}},
  \bibinfo{author}{\bibfnamefont{A.}~\bibnamefont{Fert}},
  \bibinfo{author}{\bibfnamefont{F.~N.} \bibnamefont{Van~Dau}},
  \bibinfo{author}{\bibfnamefont{F.}~\bibnamefont{Petroff}},
  \bibinfo{author}{\bibfnamefont{P.}~\bibnamefont{Etienne}},
  \bibinfo{author}{\bibfnamefont{G.}~\bibnamefont{Creuzet}},
  \bibinfo{author}{\bibfnamefont{A.}~\bibnamefont{Friederich}},
  \bibnamefont{and} \bibinfo{author}{\bibfnamefont{J.}~\bibnamefont{Chazelas}},
  \bibinfo{journal}{Physical Review Letters} \textbf{\bibinfo{volume}{61}},
  \bibinfo{pages}{2472} (\bibinfo{year}{1988}).

\bibitem[{\citenamefont{Binasch et~al.}(1989)\citenamefont{Binasch,
  Gr{\"u}nberg, Saurenbach, and Zinn}}]{binasch1989}
\bibinfo{author}{\bibfnamefont{G.}~\bibnamefont{Binasch}},
  \bibinfo{author}{\bibfnamefont{P.}~\bibnamefont{Gr{\"u}nberg}},
  \bibinfo{author}{\bibfnamefont{F.}~\bibnamefont{Saurenbach}},
  \bibnamefont{and} \bibinfo{author}{\bibfnamefont{W.}~\bibnamefont{Zinn}},
  \bibinfo{journal}{Physical Review B} \textbf{\bibinfo{volume}{39}},
  \bibinfo{pages}{4828} (\bibinfo{year}{1989}).

\bibitem[{\citenamefont{Julliere}(1975)}]{julliere1975}
\bibinfo{author}{\bibfnamefont{M.}~\bibnamefont{Julliere}},
  \bibinfo{journal}{Physics Letters A} \textbf{\bibinfo{volume}{54}},
  \bibinfo{pages}{225} (\bibinfo{year}{1975}).

\bibitem[{\citenamefont{Moodera et~al.}(1995)\citenamefont{Moodera, Kinder,
  Wong, and Meservey}}]{moodera1995}
\bibinfo{author}{\bibfnamefont{J.~S.} \bibnamefont{Moodera}},
  \bibinfo{author}{\bibfnamefont{L.~R.} \bibnamefont{Kinder}},
  \bibinfo{author}{\bibfnamefont{T.~M.} \bibnamefont{Wong}}, \bibnamefont{and}
  \bibinfo{author}{\bibfnamefont{R.}~\bibnamefont{Meservey}},
  \bibinfo{journal}{Physical Review Letters} \textbf{\bibinfo{volume}{74}},
  \bibinfo{pages}{3273} (\bibinfo{year}{1995}).

\bibitem[{\citenamefont{Dieny et~al.}(1991)\citenamefont{Dieny, Speriosu,
  Metin, Parkin, Gurney, Baumgart, and Wilhoit}}]{dieny1991}
\bibinfo{author}{\bibfnamefont{B.}~\bibnamefont{Dieny}},
  \bibinfo{author}{\bibfnamefont{V.~S.} \bibnamefont{Speriosu}},
  \bibinfo{author}{\bibfnamefont{S.}~\bibnamefont{Metin}},
  \bibinfo{author}{\bibfnamefont{S.~S.} \bibnamefont{Parkin}},
  \bibinfo{author}{\bibfnamefont{B.~A.} \bibnamefont{Gurney}},
  \bibinfo{author}{\bibfnamefont{P.}~\bibnamefont{Baumgart}}, \bibnamefont{and}
  \bibinfo{author}{\bibfnamefont{D.~R.} \bibnamefont{Wilhoit}},
  \bibinfo{journal}{Journal of Applied Physics} \textbf{\bibinfo{volume}{69}},
  \bibinfo{pages}{4774} (\bibinfo{year}{1991}).

\bibitem[{\citenamefont{Geim and Grigorieva}(2013)}]{Geim2013}
\bibinfo{author}{\bibfnamefont{A.~K.} \bibnamefont{Geim}} \bibnamefont{and}
  \bibinfo{author}{\bibfnamefont{I.~V.} \bibnamefont{Grigorieva}},
  \bibinfo{journal}{Nature} \textbf{\bibinfo{volume}{499}},
  \bibinfo{pages}{419} (\bibinfo{year}{2013}).

\bibitem[{\citenamefont{Novoselov et~al.}(2016)\citenamefont{Novoselov,
  Mishchenko, Carvalho, and Neto}}]{novoselov2016}
\bibinfo{author}{\bibfnamefont{K.}~\bibnamefont{Novoselov}},
  \bibinfo{author}{\bibfnamefont{A.}~\bibnamefont{Mishchenko}},
  \bibinfo{author}{\bibfnamefont{A.}~\bibnamefont{Carvalho}}, \bibnamefont{and}
  \bibinfo{author}{\bibfnamefont{A.~C.} \bibnamefont{Neto}},
  \bibinfo{journal}{Science} \textbf{\bibinfo{volume}{353}},
  \bibinfo{pages}{aac9439} (\bibinfo{year}{2016}).

\bibitem[{\citenamefont{Wang et~al.}(2016)\citenamefont{Wang, Du, Liu, Hu,
  Zhang, Zhang, Owen, Lu, Gan, Sengupta et~al.}}]{Wang2016}
\bibinfo{author}{\bibfnamefont{X.}~\bibnamefont{Wang}},
  \bibinfo{author}{\bibfnamefont{K.}~\bibnamefont{Du}},
  \bibinfo{author}{\bibfnamefont{Y.~Y.~F.} \bibnamefont{Liu}},
  \bibinfo{author}{\bibfnamefont{P.}~\bibnamefont{Hu}},
  \bibinfo{author}{\bibfnamefont{J.}~\bibnamefont{Zhang}},
  \bibinfo{author}{\bibfnamefont{Q.}~\bibnamefont{Zhang}},
  \bibinfo{author}{\bibfnamefont{M.~H.~S.} \bibnamefont{Owen}},
  \bibinfo{author}{\bibfnamefont{X.}~\bibnamefont{Lu}},
  \bibinfo{author}{\bibfnamefont{C.~K.} \bibnamefont{Gan}},
  \bibinfo{author}{\bibfnamefont{P.}~\bibnamefont{Sengupta}},
  \bibnamefont{et~al.}, \bibinfo{journal}{2D Materials}
  \textbf{\bibinfo{volume}{3}}, \bibinfo{pages}{031009} (\bibinfo{year}{2016}).

\bibitem[{\citenamefont{Lee et~al.}(2016)\citenamefont{Lee, Lee, Ryoo, Kang,
  Kim, Kim, Park, Park, and Cheong}}]{Lee2016}
\bibinfo{author}{\bibfnamefont{J.-U.} \bibnamefont{Lee}},
  \bibinfo{author}{\bibfnamefont{S.}~\bibnamefont{Lee}},
  \bibinfo{author}{\bibfnamefont{J.~H.} \bibnamefont{Ryoo}},
  \bibinfo{author}{\bibfnamefont{S.}~\bibnamefont{Kang}},
  \bibinfo{author}{\bibfnamefont{T.~Y.} \bibnamefont{Kim}},
  \bibinfo{author}{\bibfnamefont{P.}~\bibnamefont{Kim}},
  \bibinfo{author}{\bibfnamefont{C.-H.} \bibnamefont{Park}},
  \bibinfo{author}{\bibfnamefont{J.-G.} \bibnamefont{Park}}, \bibnamefont{and}
  \bibinfo{author}{\bibfnamefont{H.}~\bibnamefont{Cheong}},
  \bibinfo{journal}{Nano Letters} \textbf{\bibinfo{volume}{16}},
  \bibinfo{pages}{7433} (\bibinfo{year}{2016}).

\bibitem[{\citenamefont{Gong et~al.}(2017)\citenamefont{Gong, Li, Li, Ji,
  Stern, Xia, Cao, Bao, Wang, Wang et~al.}}]{Gong2017}
\bibinfo{author}{\bibfnamefont{C.}~\bibnamefont{Gong}},
  \bibinfo{author}{\bibfnamefont{L.}~\bibnamefont{Li}},
  \bibinfo{author}{\bibfnamefont{Z.}~\bibnamefont{Li}},
  \bibinfo{author}{\bibfnamefont{H.}~\bibnamefont{Ji}},
  \bibinfo{author}{\bibfnamefont{A.}~\bibnamefont{Stern}},
  \bibinfo{author}{\bibfnamefont{Y.}~\bibnamefont{Xia}},
  \bibinfo{author}{\bibfnamefont{T.}~\bibnamefont{Cao}},
  \bibinfo{author}{\bibfnamefont{W.}~\bibnamefont{Bao}},
  \bibinfo{author}{\bibfnamefont{C.}~\bibnamefont{Wang}},
  \bibinfo{author}{\bibfnamefont{Y.}~\bibnamefont{Wang}}, \bibnamefont{et~al.},
  \bibinfo{journal}{Nature} \textbf{\bibinfo{volume}{546}},
  \bibinfo{pages}{265} (\bibinfo{year}{2017}), ISSN \bibinfo{issn}{0028-0836}.

\bibitem[{\citenamefont{Huang et~al.}(2017)\citenamefont{Huang, Clark,
  Navarro-Moratalla, Klein, Cheng, Seyler, Zhong, Schmidgall, McGuire, Cobden
  et~al.}}]{Huang2017}
\bibinfo{author}{\bibfnamefont{B.}~\bibnamefont{Huang}},
  \bibinfo{author}{\bibfnamefont{G.}~\bibnamefont{Clark}},
  \bibinfo{author}{\bibfnamefont{E.}~\bibnamefont{Navarro-Moratalla}},
  \bibinfo{author}{\bibfnamefont{D.~R.} \bibnamefont{Klein}},
  \bibinfo{author}{\bibfnamefont{R.}~\bibnamefont{Cheng}},
  \bibinfo{author}{\bibfnamefont{K.~L.} \bibnamefont{Seyler}},
  \bibinfo{author}{\bibfnamefont{D.}~\bibnamefont{Zhong}},
  \bibinfo{author}{\bibfnamefont{E.}~\bibnamefont{Schmidgall}},
  \bibinfo{author}{\bibfnamefont{M.~A.} \bibnamefont{McGuire}},
  \bibinfo{author}{\bibfnamefont{D.~H.} \bibnamefont{Cobden}},
  \bibnamefont{et~al.}, \bibinfo{journal}{Nature}
  \textbf{\bibinfo{volume}{546}}, \bibinfo{pages}{270} (\bibinfo{year}{2017}).

\bibitem[{\citenamefont{Klein et~al.}(2018)\citenamefont{Klein, MacNeill, Lado,
  Soriano, Navarro-Moratalla, Watanabe, Taniguchi, Manni, Canfield,
  Fern{\'a}ndez-Rossier et~al.}}]{klein2018}
\bibinfo{author}{\bibfnamefont{D.~R.} \bibnamefont{Klein}},
  \bibinfo{author}{\bibfnamefont{D.}~\bibnamefont{MacNeill}},
  \bibinfo{author}{\bibfnamefont{J.~L.} \bibnamefont{Lado}},
  \bibinfo{author}{\bibfnamefont{D.}~\bibnamefont{Soriano}},
  \bibinfo{author}{\bibfnamefont{E.}~\bibnamefont{Navarro-Moratalla}},
  \bibinfo{author}{\bibfnamefont{K.}~\bibnamefont{Watanabe}},
  \bibinfo{author}{\bibfnamefont{T.}~\bibnamefont{Taniguchi}},
  \bibinfo{author}{\bibfnamefont{S.}~\bibnamefont{Manni}},
  \bibinfo{author}{\bibfnamefont{P.}~\bibnamefont{Canfield}},
  \bibinfo{author}{\bibfnamefont{J.}~\bibnamefont{Fern{\'a}ndez-Rossier}},
  \bibnamefont{et~al.}, \bibinfo{journal}{arXiv preprint arXiv:1801.10075}
  (\bibinfo{year}{2018}).

\bibitem[{\citenamefont{Song et~al.}(2018)\citenamefont{Song, Cai, Tu, Zhang,
  Huang, Wilson, Seyler, Zhu, Taniguchi, Watanabe et~al.}}]{song2018}
\bibinfo{author}{\bibfnamefont{T.}~\bibnamefont{Song}},
  \bibinfo{author}{\bibfnamefont{X.}~\bibnamefont{Cai}},
  \bibinfo{author}{\bibfnamefont{M.~W.-Y.} \bibnamefont{Tu}},
  \bibinfo{author}{\bibfnamefont{X.}~\bibnamefont{Zhang}},
  \bibinfo{author}{\bibfnamefont{B.}~\bibnamefont{Huang}},
  \bibinfo{author}{\bibfnamefont{N.~P.} \bibnamefont{Wilson}},
  \bibinfo{author}{\bibfnamefont{K.~L.} \bibnamefont{Seyler}},
  \bibinfo{author}{\bibfnamefont{L.}~\bibnamefont{Zhu}},
  \bibinfo{author}{\bibfnamefont{T.}~\bibnamefont{Taniguchi}},
  \bibinfo{author}{\bibfnamefont{K.}~\bibnamefont{Watanabe}},
  \bibnamefont{et~al.}, \bibinfo{journal}{arXiv preprint arXiv:1801.08679}
  (\bibinfo{year}{2018}).

\bibitem[{\citenamefont{Wang et~al.}(2018)\citenamefont{Wang,
  Guti{\'e}rrez-Lezama, Ubrig, Kroner, Taniguchi, Watanabe, Imamo{\u{g}}lu,
  Giannini, and Morpurgo}}]{wang2018}
\bibinfo{author}{\bibfnamefont{Z.}~\bibnamefont{Wang}},
  \bibinfo{author}{\bibfnamefont{I.}~\bibnamefont{Guti{\'e}rrez-Lezama}},
  \bibinfo{author}{\bibfnamefont{N.}~\bibnamefont{Ubrig}},
  \bibinfo{author}{\bibfnamefont{M.}~\bibnamefont{Kroner}},
  \bibinfo{author}{\bibfnamefont{T.}~\bibnamefont{Taniguchi}},
  \bibinfo{author}{\bibfnamefont{K.}~\bibnamefont{Watanabe}},
  \bibinfo{author}{\bibfnamefont{A.}~\bibnamefont{Imamo{\u{g}}lu}},
  \bibinfo{author}{\bibfnamefont{E.}~\bibnamefont{Giannini}}, \bibnamefont{and}
  \bibinfo{author}{\bibfnamefont{A.~F.} \bibnamefont{Morpurgo}},
  \bibinfo{journal}{arXiv preprint arXiv:1801.08188}  (\bibinfo{year}{2018}).

\bibitem[{\citenamefont{Jiang et~al.}(2018{\natexlab{a}})\citenamefont{Jiang,
  Shan, and Mak}}]{Jiang2018}
\bibinfo{author}{\bibfnamefont{S.}~\bibnamefont{Jiang}},
  \bibinfo{author}{\bibfnamefont{J.}~\bibnamefont{Shan}}, \bibnamefont{and}
  \bibinfo{author}{\bibfnamefont{K.~F.} \bibnamefont{Mak}},
  \bibinfo{journal}{Nature materials}  (\bibinfo{year}{2018}{\natexlab{a}}).

\bibitem[{\citenamefont{Huang et~al.}(2018)\citenamefont{Huang, Clark, Klein,
  MacNeill, Navarro-Moratalla, Seyler, Wilson, McGuire, Cobden, Xiao
  et~al.}}]{Huang2018}
\bibinfo{author}{\bibfnamefont{B.}~\bibnamefont{Huang}},
  \bibinfo{author}{\bibfnamefont{G.}~\bibnamefont{Clark}},
  \bibinfo{author}{\bibfnamefont{D.~R.} \bibnamefont{Klein}},
  \bibinfo{author}{\bibfnamefont{D.}~\bibnamefont{MacNeill}},
  \bibinfo{author}{\bibfnamefont{E.}~\bibnamefont{Navarro-Moratalla}},
  \bibinfo{author}{\bibfnamefont{K.~L.} \bibnamefont{Seyler}},
  \bibinfo{author}{\bibfnamefont{N.}~\bibnamefont{Wilson}},
  \bibinfo{author}{\bibfnamefont{M.~A.} \bibnamefont{McGuire}},
  \bibinfo{author}{\bibfnamefont{D.~H.} \bibnamefont{Cobden}},
  \bibinfo{author}{\bibfnamefont{D.}~\bibnamefont{Xiao}}, \bibnamefont{et~al.},
  \bibinfo{journal}{arXiv:1802.06979v2 [cond-mat.mes-hall]}
  (\bibinfo{year}{2018}).

\bibitem[{\citenamefont{Jiang et~al.}(2018{\natexlab{b}})\citenamefont{Jiang,
  Li, Wang, Mak, and Shan}}]{Jiang2018b}
\bibinfo{author}{\bibfnamefont{S.}~\bibnamefont{Jiang}},
  \bibinfo{author}{\bibfnamefont{L.}~\bibnamefont{Li}},
  \bibinfo{author}{\bibfnamefont{Z.}~\bibnamefont{Wang}},
  \bibinfo{author}{\bibfnamefont{K.~F.} \bibnamefont{Mak}}, \bibnamefont{and}
  \bibinfo{author}{\bibfnamefont{J.}~\bibnamefont{Shan}},
  \bibinfo{journal}{arXiv:1802.07355v1}  (\bibinfo{year}{2018}{\natexlab{b}}).

\bibitem[{\citenamefont{Tombros et~al.}(2007)\citenamefont{Tombros, Jozsa,
  Popinciuc, Jonkman, and Van~Wees}}]{tombros2007}
\bibinfo{author}{\bibfnamefont{N.}~\bibnamefont{Tombros}},
  \bibinfo{author}{\bibfnamefont{C.}~\bibnamefont{Jozsa}},
  \bibinfo{author}{\bibfnamefont{M.}~\bibnamefont{Popinciuc}},
  \bibinfo{author}{\bibfnamefont{H.~T.} \bibnamefont{Jonkman}},
  \bibnamefont{and} \bibinfo{author}{\bibfnamefont{B.~J.}
  \bibnamefont{Van~Wees}}, \bibinfo{journal}{Nature}
  \textbf{\bibinfo{volume}{448}}, \bibinfo{pages}{571} (\bibinfo{year}{2007}).

\bibitem[{\citenamefont{Han et~al.}(2014)\citenamefont{Han, Kawakami, Gmitra,
  and Fabian}}]{han2014}
\bibinfo{author}{\bibfnamefont{W.}~\bibnamefont{Han}},
  \bibinfo{author}{\bibfnamefont{R.~K.} \bibnamefont{Kawakami}},
  \bibinfo{author}{\bibfnamefont{M.}~\bibnamefont{Gmitra}}, \bibnamefont{and}
  \bibinfo{author}{\bibfnamefont{J.}~\bibnamefont{Fabian}},
  \bibinfo{journal}{Nature nanotechnology} \textbf{\bibinfo{volume}{9}},
  \bibinfo{pages}{794} (\bibinfo{year}{2014}).

\bibitem[{\citenamefont{Miao et~al.}(2009)\citenamefont{Miao, M\"uller, and
  Moodera}}]{Miao2009}
\bibinfo{author}{\bibfnamefont{G.-X.} \bibnamefont{Miao}},
  \bibinfo{author}{\bibfnamefont{M.}~\bibnamefont{M\"uller}}, \bibnamefont{and}
  \bibinfo{author}{\bibfnamefont{J.~S.} \bibnamefont{Moodera}},
  \bibinfo{journal}{Phys. Rev. Lett.} \textbf{\bibinfo{volume}{102}},
  \bibinfo{pages}{076601} (\bibinfo{year}{2009}).

\bibitem[{\citenamefont{Guinea et~al.}(2006)\citenamefont{Guinea, Neto, and
  Peres}}]{guinea2006}
\bibinfo{author}{\bibfnamefont{F.}~\bibnamefont{Guinea}},
  \bibinfo{author}{\bibfnamefont{A.~C.} \bibnamefont{Neto}}, \bibnamefont{and}
  \bibinfo{author}{\bibfnamefont{N.}~\bibnamefont{Peres}},
  \bibinfo{journal}{Physical Review B} \textbf{\bibinfo{volume}{73}},
  \bibinfo{pages}{245426} (\bibinfo{year}{2006}).

\bibitem[{\citenamefont{Phong et~al.}(2017)\citenamefont{Phong, Walet, and
  Guinea}}]{phong2017}
\bibinfo{author}{\bibfnamefont{V.~T.} \bibnamefont{Phong}},
  \bibinfo{author}{\bibfnamefont{N.~R.} \bibnamefont{Walet}}, \bibnamefont{and}
  \bibinfo{author}{\bibfnamefont{F.}~\bibnamefont{Guinea}},
  \bibinfo{journal}{2D Materials} \textbf{\bibinfo{volume}{5}},
  \bibinfo{pages}{014004} (\bibinfo{year}{2017}).

\bibitem[{\citenamefont{Yang et~al.}(2013)\citenamefont{Yang, Hallal, Terrade,
  Waintal, Roche, and Chshiev}}]{yang2013}
\bibinfo{author}{\bibfnamefont{H.}~\bibnamefont{Yang}},
  \bibinfo{author}{\bibfnamefont{A.}~\bibnamefont{Hallal}},
  \bibinfo{author}{\bibfnamefont{D.}~\bibnamefont{Terrade}},
  \bibinfo{author}{\bibfnamefont{X.}~\bibnamefont{Waintal}},
  \bibinfo{author}{\bibfnamefont{S.}~\bibnamefont{Roche}}, \bibnamefont{and}
  \bibinfo{author}{\bibfnamefont{M.}~\bibnamefont{Chshiev}},
  \bibinfo{journal}{Physical Review Letters} \textbf{\bibinfo{volume}{110}},
  \bibinfo{pages}{046603} (\bibinfo{year}{2013}).

\bibitem[{\citenamefont{Hallal et~al.}(2017)\citenamefont{Hallal, Ibrahim,
  Yang, Roche, and Chshiev}}]{hallal2017}
\bibinfo{author}{\bibfnamefont{A.}~\bibnamefont{Hallal}},
  \bibinfo{author}{\bibfnamefont{F.}~\bibnamefont{Ibrahim}},
  \bibinfo{author}{\bibfnamefont{H.}~\bibnamefont{Yang}},
  \bibinfo{author}{\bibfnamefont{S.}~\bibnamefont{Roche}}, \bibnamefont{and}
  \bibinfo{author}{\bibfnamefont{M.}~\bibnamefont{Chshiev}},
  \bibinfo{journal}{2D Materials} \textbf{\bibinfo{volume}{4}},
  \bibinfo{pages}{025074} (\bibinfo{year}{2017}).

\bibitem[{\citenamefont{Ohta et~al.}(2006)\citenamefont{Ohta, Bostwick,
  Seyller, Horn, and Rotenberg}}]{ohta2006}
\bibinfo{author}{\bibfnamefont{T.}~\bibnamefont{Ohta}},
  \bibinfo{author}{\bibfnamefont{A.}~\bibnamefont{Bostwick}},
  \bibinfo{author}{\bibfnamefont{T.}~\bibnamefont{Seyller}},
  \bibinfo{author}{\bibfnamefont{K.}~\bibnamefont{Horn}}, \bibnamefont{and}
  \bibinfo{author}{\bibfnamefont{E.}~\bibnamefont{Rotenberg}},
  \bibinfo{journal}{Science} \textbf{\bibinfo{volume}{313}},
  \bibinfo{pages}{951} (\bibinfo{year}{2006}).

\bibitem[{\citenamefont{McCann}(2006)}]{mccann2006}
\bibinfo{author}{\bibfnamefont{E.}~\bibnamefont{McCann}},
  \bibinfo{journal}{Phys. Rev. B} \textbf{\bibinfo{volume}{74}},
  \bibinfo{pages}{161403} (\bibinfo{year}{2006}).

\bibitem[{\citenamefont{Fern{\'a}ndez-Rossier}(2008)}]{JFR2008}
\bibinfo{author}{\bibfnamefont{J.}~\bibnamefont{Fern{\'a}ndez-Rossier}},
  \bibinfo{journal}{Physical Review B} \textbf{\bibinfo{volume}{77}},
  \bibinfo{pages}{075430} (\bibinfo{year}{2008}).

\bibitem[{\citenamefont{Martin et~al.}(2008)\citenamefont{Martin, Blanter, and
  Morpurgo}}]{martin2008}
\bibinfo{author}{\bibfnamefont{I.}~\bibnamefont{Martin}},
  \bibinfo{author}{\bibfnamefont{Y.~M.} \bibnamefont{Blanter}},
  \bibnamefont{and} \bibinfo{author}{\bibfnamefont{A.}~\bibnamefont{Morpurgo}},
  \bibinfo{journal}{Physical Review Letters} \textbf{\bibinfo{volume}{100}},
  \bibinfo{pages}{036804} (\bibinfo{year}{2008}).

\bibitem[{\citenamefont{San-Jose et~al.}(2009)\citenamefont{San-Jose, Prada,
  McCann, and Schomerus}}]{pablo2009}
\bibinfo{author}{\bibfnamefont{P.}~\bibnamefont{San-Jose}},
  \bibinfo{author}{\bibfnamefont{E.}~\bibnamefont{Prada}},
  \bibinfo{author}{\bibfnamefont{E.}~\bibnamefont{McCann}}, \bibnamefont{and}
  \bibinfo{author}{\bibfnamefont{H.}~\bibnamefont{Schomerus}},
  \bibinfo{journal}{Physical Review Letters} \textbf{\bibinfo{volume}{102}},
  \bibinfo{pages}{247204} (\bibinfo{year}{2009}).

\bibitem[{\citenamefont{Lado et~al.}(2015)\citenamefont{Lado,
  Garc{\'\i}a-Mart{\'\i}nez, and Fern{\'a}ndez-Rossier}}]{Lado2015}
\bibinfo{author}{\bibfnamefont{J.~L.} \bibnamefont{Lado}},
  \bibinfo{author}{\bibfnamefont{N.}~\bibnamefont{Garc{\'\i}a-Mart{\'\i}nez}},
  \bibnamefont{and}
  \bibinfo{author}{\bibfnamefont{J.}~\bibnamefont{Fern{\'a}ndez-Rossier}},
  \bibinfo{journal}{Synthetic Metals} \textbf{\bibinfo{volume}{210}},
  \bibinfo{pages}{56} (\bibinfo{year}{2015}).

\bibitem[{\citenamefont{Zhong et~al.}(2017)\citenamefont{Zhong, Seyler,
  Linpeng, Cheng, Sivadas, Huang, Schmidgall, Taniguchi, Watanabe, McGuire
  et~al.}}]{zhong2017}
\bibinfo{author}{\bibfnamefont{D.}~\bibnamefont{Zhong}},
  \bibinfo{author}{\bibfnamefont{K.~L.} \bibnamefont{Seyler}},
  \bibinfo{author}{\bibfnamefont{X.}~\bibnamefont{Linpeng}},
  \bibinfo{author}{\bibfnamefont{R.}~\bibnamefont{Cheng}},
  \bibinfo{author}{\bibfnamefont{N.}~\bibnamefont{Sivadas}},
  \bibinfo{author}{\bibfnamefont{B.}~\bibnamefont{Huang}},
  \bibinfo{author}{\bibfnamefont{E.}~\bibnamefont{Schmidgall}},
  \bibinfo{author}{\bibfnamefont{T.}~\bibnamefont{Taniguchi}},
  \bibinfo{author}{\bibfnamefont{K.}~\bibnamefont{Watanabe}},
  \bibinfo{author}{\bibfnamefont{M.~A.} \bibnamefont{McGuire}},
  \bibnamefont{et~al.}, \bibinfo{journal}{Science Advances}
  \textbf{\bibinfo{volume}{3}}, \bibinfo{pages}{e1603113}
  (\bibinfo{year}{2017}).

\bibitem[{\citenamefont{Giannozzi et~al.}(2009)\citenamefont{Giannozzi, Baroni,
  Bonini, Calandra, Car, Cavazzoni, Ceresoli, Chiarotti, Cococcioni, Dabo
  et~al.}}]{Giannozzi_JPhysConMat_2009}
\bibinfo{author}{\bibfnamefont{P.}~\bibnamefont{Giannozzi}},
  \bibinfo{author}{\bibfnamefont{S.}~\bibnamefont{Baroni}},
  \bibinfo{author}{\bibfnamefont{N.}~\bibnamefont{Bonini}},
  \bibinfo{author}{\bibfnamefont{M.}~\bibnamefont{Calandra}},
  \bibinfo{author}{\bibfnamefont{R.}~\bibnamefont{Car}},
  \bibinfo{author}{\bibfnamefont{C.}~\bibnamefont{Cavazzoni}},
  \bibinfo{author}{\bibfnamefont{D.}~\bibnamefont{Ceresoli}},
  \bibinfo{author}{\bibfnamefont{G.~L.} \bibnamefont{Chiarotti}},
  \bibinfo{author}{\bibfnamefont{M.}~\bibnamefont{Cococcioni}},
  \bibinfo{author}{\bibfnamefont{I.}~\bibnamefont{Dabo}}, \bibnamefont{et~al.},
  \bibinfo{journal}{J. Phys.: Condens. Mat.} \textbf{\bibinfo{volume}{21}},
  \bibinfo{pages}{395502} (\bibinfo{year}{2009}).

\bibitem[{\citenamefont{Perdew et~al.}(1996)\citenamefont{Perdew, Ernzerhof,
  and Burke}}]{Perdew_JCP_1996}
\bibinfo{author}{\bibfnamefont{J.~P.} \bibnamefont{Perdew}},
  \bibinfo{author}{\bibfnamefont{M.}~\bibnamefont{Ernzerhof}},
  \bibnamefont{and} \bibinfo{author}{\bibfnamefont{K.}~\bibnamefont{Burke}},
  \bibinfo{journal}{J. Chem. Phys.} \textbf{\bibinfo{volume}{105}},
  \bibinfo{pages}{9982} (\bibinfo{year}{1996}).

\bibitem[{\citenamefont{Bl\"ochl}(1994)}]{PhysRevB.50.17953}
\bibinfo{author}{\bibfnamefont{P.~E.} \bibnamefont{Bl\"ochl}},
  \bibinfo{journal}{Phys. Rev. B} \textbf{\bibinfo{volume}{50}},
  \bibinfo{pages}{17953} (\bibinfo{year}{1994}).

\bibitem[{\citenamefont{Kucukbenli et~al.}(2014)\citenamefont{Kucukbenli,
  Monni, Adetunji, Ge, Adebayo, Marzari, De~Gironcoli, and
  Corso}}]{kucukbenli2014projector}
\bibinfo{author}{\bibfnamefont{E.}~\bibnamefont{Kucukbenli}},
  \bibinfo{author}{\bibfnamefont{M.}~\bibnamefont{Monni}},
  \bibinfo{author}{\bibfnamefont{B.}~\bibnamefont{Adetunji}},
  \bibinfo{author}{\bibfnamefont{X.}~\bibnamefont{Ge}},
  \bibinfo{author}{\bibfnamefont{G.}~\bibnamefont{Adebayo}},
  \bibinfo{author}{\bibfnamefont{N.}~\bibnamefont{Marzari}},
  \bibinfo{author}{\bibfnamefont{S.}~\bibnamefont{De~Gironcoli}},
  \bibnamefont{and} \bibinfo{author}{\bibfnamefont{A.~D.} \bibnamefont{Corso}},
  \bibinfo{journal}{arXiv preprint arXiv:1404.3015}  (\bibinfo{year}{2014}).

\bibitem[{\citenamefont{Grimme}(2006)}]{Grimme_JCompChem_2006}
\bibinfo{author}{\bibfnamefont{S.}~\bibnamefont{Grimme}}, \bibinfo{journal}{J.
  Comput. Chem.} \textbf{\bibinfo{volume}{27}}, \bibinfo{pages}{1787}
  (\bibinfo{year}{2006}).

\bibitem[{\citenamefont{Lado and Fern{\'a}ndez-Rossier}(2017)}]{Lado2017}
\bibinfo{author}{\bibfnamefont{J.~L.} \bibnamefont{Lado}} \bibnamefont{and}
  \bibinfo{author}{\bibfnamefont{J.}~\bibnamefont{Fern{\'a}ndez-Rossier}},
  \bibinfo{journal}{2D Materials} \textbf{\bibinfo{volume}{4}},
  \bibinfo{pages}{035002} (\bibinfo{year}{2017}).

\bibitem[{\citenamefont{Zhang et~al.}(2017)\citenamefont{Zhang, Zhao, Zhou,
  Xue, Ma, and Yang}}]{zhang2017}
\bibinfo{author}{\bibfnamefont{J.}~\bibnamefont{Zhang}},
  \bibinfo{author}{\bibfnamefont{B.}~\bibnamefont{Zhao}},
  \bibinfo{author}{\bibfnamefont{T.}~\bibnamefont{Zhou}},
  \bibinfo{author}{\bibfnamefont{Y.}~\bibnamefont{Xue}},
  \bibinfo{author}{\bibfnamefont{C.}~\bibnamefont{Ma}}, \bibnamefont{and}
  \bibinfo{author}{\bibfnamefont{Z.}~\bibnamefont{Yang}},
  \bibinfo{journal}{arXiv preprint arXiv:1710.06324}  (\bibinfo{year}{2017}).

\bibitem[{\citenamefont{Young et~al.}(2014)\citenamefont{Young,
  Sanchez-Yamagishi, Hunt, Choi, Watanabe, Taniguchi, Ashoori, and
  Jarillo-Herrero}}]{Young2014}
\bibinfo{author}{\bibfnamefont{A.}~\bibnamefont{Young}},
  \bibinfo{author}{\bibfnamefont{J.}~\bibnamefont{Sanchez-Yamagishi}},
  \bibinfo{author}{\bibfnamefont{B.}~\bibnamefont{Hunt}},
  \bibinfo{author}{\bibfnamefont{S.}~\bibnamefont{Choi}},
  \bibinfo{author}{\bibfnamefont{K.}~\bibnamefont{Watanabe}},
  \bibinfo{author}{\bibfnamefont{T.}~\bibnamefont{Taniguchi}},
  \bibinfo{author}{\bibfnamefont{R.}~\bibnamefont{Ashoori}}, \bibnamefont{and}
  \bibinfo{author}{\bibfnamefont{P.}~\bibnamefont{Jarillo-Herrero}},
  \bibinfo{journal}{Nature} \textbf{\bibinfo{volume}{505}},
  \bibinfo{pages}{528} (\bibinfo{year}{2014}).

\bibitem[{\citenamefont{Li et~al.}(2013)\citenamefont{Li, Roschewsky, Assaf,
  Eich, Epstein-Martin, Heiman, M{\"u}nzenberg, and Moodera}}]{li2013}
\bibinfo{author}{\bibfnamefont{B.}~\bibnamefont{Li}},
  \bibinfo{author}{\bibfnamefont{N.}~\bibnamefont{Roschewsky}},
  \bibinfo{author}{\bibfnamefont{B.~A.} \bibnamefont{Assaf}},
  \bibinfo{author}{\bibfnamefont{M.}~\bibnamefont{Eich}},
  \bibinfo{author}{\bibfnamefont{M.}~\bibnamefont{Epstein-Martin}},
  \bibinfo{author}{\bibfnamefont{D.}~\bibnamefont{Heiman}},
  \bibinfo{author}{\bibfnamefont{M.}~\bibnamefont{M{\"u}nzenberg}},
  \bibnamefont{and} \bibinfo{author}{\bibfnamefont{J.~S.}
  \bibnamefont{Moodera}}, \bibinfo{journal}{Physical Review Letters}
  \textbf{\bibinfo{volume}{110}}, \bibinfo{pages}{097001}
  (\bibinfo{year}{2013}).

\end{thebibliography}

\end{document}